\documentclass[superscriptaddress,iop,apj]{emulateapj}
\usepackage{amssymb}
\usepackage{amsfonts}
\usepackage{amsmath}
\usepackage{bm}
\usepackage[normalem]{ulem}
\usepackage{multirow}
\usepackage{subfigure}
\usepackage{color}
\usepackage{calc}
\usepackage{amsmath,amssymb,graphicx}
\usepackage{amssymb,amsmath}
\usepackage{tensor}
\usepackage{bm}
\usepackage{microtype}
\usepackage{booktabs}
\usepackage{times}
\usepackage[varg]{txfonts}
\usepackage[colorlinks, pdfborder={0 0 0}]{hyperref}
\usepackage{subfigure}
\usepackage{lipsum}
\usepackage{breakurl}
\usepackage{graphicx}
\usepackage{pbox}
\usepackage{threeparttable}                                                     

\newcommand{\saxj}{\mbox{SAX J1808.4$-$3658}}
\newcommand{\igr}{\mbox{IGR J0029+5934}}

\newcommand{\chandra}{\textit{Chandra}}

\newcommand{\msun}{\,M_{\odot}}

\newcommand{\ergs}{\rm\,erg\,s^{-1}}

\begin{document}

\title{Radio pulse search and X-Ray monitoring of \saxj\,: What Causes its Orbital Evolution?} 
\shorttitle{17 years of evolution \saxj}
\shortauthors{Patruno et~al.}

\author{
  Alessandro Patruno\altaffilmark{1,2},
  Amruta Jaodand\altaffilmark{2,3}
  Lucien Kuiper\altaffilmark{4},
  Peter Bult\altaffilmark{3,5},
  Jason Hessels\altaffilmark{2,3},
  Christian Knigge\altaffilmark{6}
  Andrew R. King\altaffilmark{7,1,3},
Rudy Wijnands\altaffilmark{3},
Michiel van der Klis\altaffilmark{3}}
\altaffiltext{1}{Leiden Observatory, Leiden University,
              Neils Bohrweg 2, 2333 CA, Leiden, The Netherlands}
   \altaffiltext{2}{ASTRON, the Netherlands Institute for Radio Astronomy, Postbus 2, 7900 AA, Dwingeloo, the Netherlands}  
\altaffiltext{3}{Anton Pannekoek Institute for Astronomy, University of Amsterdam, Science\
 Park 904, 1098 XH, Amsterdam, The Netherlands}  
\altaffiltext{4}{SRON-National Institute for Space Research, Sorbonnelaan 2, NL-3584 CA Utrecht, the Netherlands}
\altaffiltext{5}{Astrophysics Science Division, NASA Goddard Space Flight Center
 Greenbelt, MD 20771, USA}
\altaffiltext{6}{University of Southampton, School of Physics and Astronomy, Southampton SO17 1BJ, UK}
\altaffiltext{7}{Theoretical Astrophysics Group, Department of Physics and Astronomy, University of Leicester, Leicester LE1 7RH, UK}
\begin{abstract}

The accreting millisecond X-ray pulsar (AMXP) \saxj\, shows a
peculiar orbital evolution that proceeds at a much faster pace than
predicted by conservative binary evolution models.  It is important to
identify the underlying mechanism responsible for this behavior
because it can help to understand how this system evolves. It has also been
suggested that, when in quiescence, SAX J1808.4-3658 turns on as a
radio pulsar, a circumstance that might provide a link between
AMXPs and black-widow radio pulsars.  In this
work we report the results of a deep radio pulsation search at 2 GHz
using the Green Bank Telescope in August 2014 and an X-ray monitoring
of the 2015 outburst with Chandra, Swift, and INTEGRAL. In particular,
we present the X-ray timing analysis of a 30-ks Chandra observation
executed during the 2015 outburst. We detect no radio pulsations, and
place the strongest limit to date on the pulsed radio flux density of
any AMXP. We also find that the orbit of SAX J1808.4-3658
continues evolving at a fast pace and we compare it to the bhevior of
other accreting and non-accreting binaries. We discuss two
scenarios: either the neutron star has a large moment of
inertia ($I>1.7x10^{45}\rm\,g\,cm^2$) and is ablating the donor (by using its
spin-down power) thus generating mass-loss with an efficiency of 40\%
or the donor star is undergoing quasi-cyclic variations due to a
varying mass-quadrupole induced by either a strong (1 kG) field or by
some unidentified mechanism probably linked to irradiation.

\end{abstract}



\section{Introduction}

The accreting millisecond X-ray pulsar (AMXP) \saxj\, is an accreting
neutron star located at a distance of 3.5~kpc~\citep{gal06} that is
spinning at 401 Hz \citep{wij98} and orbiting its 0.05-0.08$\msun$
companion in 2.01 hours~\citep{cha98,del08,wan13}.  This source was
discovered by \textit{BeppoSAX} in 1996 \citep{int98} and is the best
studied AMXP of all 18 known members (see \citealt{pat12r} for a
review). It has shown eight outbursts so far, observed with a
recurrence time of approximately 3--4 years. The high time and/or
spectral resolution of X-ray telescopes like \textit{RXTE},
\textit{XMM-Newton}, \textit{INTEGRAL}, \textit{Chandra},
\textit{Swift} and \textit{Suzaku} has allowed a thorough study of the
pulsations (see e.g., \citealt{har08,bur09,pat12}), its aperiodic
timing variability \citep{wij01,wij03,pat09c,bul15} and X-ray spectral
properties \citep{gie02,pou03,cac09,pap09,pat09e}.

The coherent timing of the pulsations has revealed the lack of a
strong spin up during the outbursts~\citep{har08,har09} and a constant
spin-down in quiescence that is compatible with magnetic dipole energy
loss (surface magnetic field $B\approx 10^{8}$ G,
see~\citealt{har08,dis08,har09,pat12}).  There is also indirect
observational evidence that \saxj\, turns on as a radio pulsar during
quiescence, although no radio pulsations have been detected so
far~\citep{hom01, bur03,cam04}.  Indeed, the optical counterpart of
\saxj\, is over-luminous with respect to a non-irradiated brown-dwarf
model \citep{bil01} during this phase. A source of irradiation is
required to explain this behavior, but the feeble X-ray
irradiation coming from the accretion disk/neutron star surface during
quiescence~\citep{hom01,hei09} cannot account for the donor
luminosity. It has been speculated that a pulsar wind impinging on the
donor surface~\citep{bur03,cam04} might be responsible for the
observed excess luminosity\footnote{Although the requirement of a
  pulsar wind is often mentioned as indirect evidence for a radio
  pulsar turning on, it is also possible that a pulsar wind is active
  without the radio pulsar mechanism operating in the system (or at
  least without radio pulsations being observable; they may be
  obscured by intra-binary material; see e.g., \citealt{jao16})}.
Optical modulation at the orbital period is now well established in
black-widow and redback radio pulsar systems \citep{bre13} and
something similar has been found for \saxj\, too during quiescence
\citep{del08,wan13}. In recent works, \citet{xin15} and \citet{deo16}
have identified a possible gamma-ray counterpart of \saxj\, and
spectral modeling of the \textit{FERMI}/LAT data imply that (if the
counterpart is confirmed) about 30\% of the spin-down energy is
transformed into gamma-rays, providing further evidence in favor of
this scenario, although no gamma-ray pulsations have been found so far.

The orbital evolution of \saxj\, shows an increase of the orbital
period on a relatively short timescale of
${\sim}70$~Myr~\citep{har08,dis08} and an acceleration of the rate of
expansion of the orbit up to 2011~\citep{pat12}. The driving mechanism
for the evolution of a binary with a ${\sim}2$ hr orbital period like
\saxj\, is expected to be angular momentum loss due to gravitational
wave emission and the expected orbital evolution timescale in this
case is ${\sim}4$ Gyr~\citep{har08}, which is almost two orders of
magnitude longer than the observed one. This behavior might require
that an additional mechanism has a strong influence on the orbit,
beside the gravitational wave emission.

Anomalously fast orbital evolution is not unique to \saxj\,. 
Several other low-mass X-ray binaries (LMXBs),
comprising both neutron star and black hole accretors, are also
observed to show a faster evolution than expected (see
Section~\ref{sec:discussion} for an in-depth discussion). There is
also a similar behavior in many (non-accreting) binary radio
millisecond pulsars with orbital parameters similar to \saxj\,, known
as ``black-widows'' (BWs), where the rotational power emitted in form
of wind and radiation by the pulsar is impinging and ablating the
semi-degenerate\footnote{In this paper we use the term
  ``semi-degenerate'' instead of brown-dwarf because the mass transfer
  process alters significantly the internal structure of these stars
  which are quite different from isolated brown dwarfs~\citep{tau11}.}
donor companion \citep{nic00,dor01,laz11}.  These systems have a
companion star with a typical mass of $\lesssim\,0.1\msun$ and several
(but not all) of them have orbital periods of about 1--3 hours. Most
BWs have very short binary evolution timescales too, orders of
magnitude shorter than the expected theoretical values for their
secular evolution. These short timescales variations are believed to
reflect some short-term effects rather than the secular evolution of
the binary.

For \saxj\, \citet{dis08} proposed a scenario in which the pulsar
wind, powered by the rotational spin-down of the neutron star in
quiescence, causes the ejection of the gas flowing through the inner
Lagrangian point $L_1$ (radio-ejection scenario; see also
\citealt{bur01,bur09}).  \citet{har08,har09} and \citet{pat12}
proposed an alternative mechanism where the binary evolution
is not necessarily driven by the matter expulsion, but it is rather a
quasi-stochastic process due to the development of a significant mass
quadrupole in the donor star that results in a coupling between the
donor spin and the orbital period of the binary. \citet{pat12} in
particular showed that the so-called Applegate mechanism
\citep{app92,app94}, which assumes that a mass quadrupole develops in
the donor star due to quasi-periodic magnetic cycles, seems to be a
promising candidate surviving the observational scrutiny. However,
there is not yet any conclusive evidence about the exact operating
mechanism behind the orbital evolution of \saxj\,.

Motivated by these facts we have conducted a deep radio pulse search
of \saxj\, during quiescence and an X-ray monitoring during the last
2015 outburst. The deep radio pulse search was done with the Greenbank Radio Telescope in August 2014 (during quiescence)
to answer the question: does \saxj\, really turn on as a radio pulsar
during quiescence? A detection could help to solve two problems: the first
is that it would allow a continuous monitoring of the orbital evolution
not only during outbursts but also during quiescence. The second is that it can allow the precise
measurement of the spin-down power of the pulsar which might play a
fundamental role in the ablation of the companion. 
The X-ray monitoring campaign was made
with the \textit{Swift} X-Ray Telescope (XRT), the \textit{INTEGRAL}
and \textit{Chandra} telescopes during the 2015 outburst.  We
used the \textit{Chandra} telescope to monitor the pulsations of the
source and track the long term orbital evolution of \saxj\,. The
question we seek to answer is whether the determination of a new
orbital solution that includes also the 2015 outburst can provide
hints on the exact mechanism behind the rapid orbital evolution of the
system.

\section{X-ray Observation and Data Reduction}

To construct the outburst light curve we analyzed all pointed {\it
  Swift}/XRT observations taken between April 1st, 2015 (MJD 57113)
and August 26, 2015 (MJD~57260). During this period 62 observations were
taken (ProgamIDs 33737, 33737 and 33801) in either window-timing mode
(1.76-ms resolution) or photon-counting mode (2.5-s). We extracted
X-ray count rates averaged per spacecraft orbit using the online {\it
  Swift}/XRT data products generator \citep{Evans2007}. To estimate
the count rate to flux conversion ratio we also used this tool to
create and fit energy spectra \citep{Evans2009}, using the $0.3-10$
keV energy and the default event grades. One Type I (i.e.,
thermonuclear) X-ray burst was detected and excluded from our analysis
(see Figure~\ref{fig:lc}).

We also used data recorded with the \textit{INTEGRAL} spacecraft,
which carries three high-energy instruments: a high-angular resolution
imager IBIS, a high-energy resolution spectrometer SPI and an X-ray
monitoring instrument JEM-X. These instruments are equipped with coded
aperture masks enabling image reconstruction in the hard X-ray/soft
$\gamma$-ray band.  Driven by sensitivity considerations, we used only
data from the \textit{INTEGRAL} Soft Gamma-Ray Imager
ISGRI~\citep{leb03}, the upper detector layer of IBIS~\citep{ube03},
sensitive to photons with energies in the range ${\sim}20$ keV -- 1
MeV (effectively ${\sim}300$ keV).
Typical
integration times are in the range 1800--3600
s.
We used the imaging software
tools~\citep{gol03} of the Offline Scientific Analysis (OSA) package
version 10.1 distributed by the \textit{INTEGRAL} Science Data Centre
(ISDC, see e.g. \citealt{cou03}).
\textit{INTEGRAL} had the
Galactic center/bulge region often in its field of view during the
April 2015 outburst of \saxj\, and the initial part of the outburst
was properly sampled. The lightcurve for the 20-100 keV range during
the early phase of outburst is shown in
Figure~\ref{fig:integral}. Each data point represents the averaged
count rate of a combination of typical 3--5 Science Windows. The onset
of the outburst, somewhat before MJD 57121, is clearly visible in this
figure.

For the timing analysis we use \chandra\, data taken with the High
Resolution Camera (HRC) with the HRC-S detector operating in timing
mode. The observation started on May 24, 2015 at 22:23:18 UT
(MJD~57166.9) and ended on May 25, 2015 at 07:15:30 UT (MJD~57167.3)
for a total exposure time of 29.6-ks. In this configuration the data
are collected with a time resolution of 16 $\mu$s and very limited energy
resolution. The \chandra\, data were processed with the CIAO software
(v 4.6) and were barycentered with the \textit{faxbary} tool by using
the most precise optical position available \citep{har08} and the JPL
DE405 solar system ephemeris.  The pulse profiles are generated by
folding the data in stretches of $\sim2000$s in pulse profiles composed
by 32 bins.

The folding procedure uses the ephemeris reported in
\citet{pat12} and extrapolates the solution to the time of the
\chandra\, observation.  Given the low signal-to-noise of the
observations we only measured the time of arrivals of the fundamental
pulse frequency ($\nu$) which also prevents that pulse shape variability affects the
fiducial point defining the pulse time of arrival (ToA; see
\citealt{har08} and \citealt{pat10} for details of the procedure).  To follow the
evolution of the orbit and the pulsar spin we fit the ToAs with the
software TEMPO2 \citep{hob06} and after obtaining a new ephemeris we
re-fold the data and repeat the procedure until convergence of the
solution.

We also created power density spectra of the \chandra\, data.  No
background subtraction was applied to the data before calculating the
power spectra. The Poissonian noise level was measured by taking the
average power between 3000 and 4000 Hz, a region dominated by counting
statistics noise alone. After obtaining the mean Poissonian value we
subtracted it from the power spectra.

We used 128-s long segments to calculate the power spectra so that our
frequency boundaries are $1/128$ Hz
and 4096 Hz.  The powers were normalized in the rms
normalization \citep{van95} which gives the power density in units of
$\rm\,(rms/mean)^2\,Hz^{-1}$. We define
the fractional rms amplitude between the frequencies $\nu_1$ and $\nu_2$ as:
\begin{equation}
\rm\,rms\it\, = \left[\int^{\nu_2}_{\nu_1}P(\nu)d\nu\right]^{\rm\,1/2}
\end{equation} 
and calculate the errors from the dispersion of the data points in the power
spectra.

\section{Radio Observations}
\saxj\, was observed on two different occasions, 2014 August 9 and 22
(MJD~$56878$ and MJD~$56891$, respectively), using the 110-m Robert
C. Byrd Green Bank Telescope (GBT), in West Virginia.  During this
period, it was known to be in X-ray quiescence, with the previous
outburst having ended in 2011, and the next outburst starting 2015
April. The data were recorded using the Green Bank Ultimate Pulsar
Processing Instrument (GUPPI) backend.  This combination of GBT with
GUPPI provides a high sensitivity to faint millisecond radio
pulsations -- arguably the deepest search that can be done with
current radio telescopes, given that the source is well outside the
Arecibo-visible declination range.

The distance to the source (${\sim}3$\,kpc) suggests a relatively high
expected dispersion measure (DM$ \gtrsim 100$\,pc\,cm$^{-3}$, based on
the NE2001 model of \citealt{CL1:2002,CL2:2003}).  Furthermore, there is the potential
for radio eclipses from intra-binary material -- in analogy with the
rotation-powered black widow and redback millisecond pulsar systems,
where the eclipse duration is typically longer at lower radio
frequencies, e.g. \citet{AKH:2013}.  Hence, our observations were
conducted at a relatively high central observing frequency of $2$\,GHz
to mitigate these effects, while still maintaining sensitivity to the
typically steep spectra of radio pulsars ($f^{-1.4}$, where $f$ is the
frequency of the electromagnetic radiation; see \cite{BLV:2013}).
GUPPI provided $800$\,MHz of bandwidth, with 61.44\,$\mu$s samples and
0.391\,MHz channels recorded as 8-bit samples in {\tt psrfits} format.
The orthogonal polarizations were summed in quadrature, providing only
total intensity.  We acquired 60/30-min integration on 2014 August 9
and 22, respectively.  The observational setup and offline data
analysis (see \S\ref{section:DA}) were tested using the millisecond pulsar
PSR~J$1824$--$2452$A (M28A).
Radio pulsations from
  M28A were easily recovered at the known pulsar spin frequency of
  327.4 Hz and $\rm\,DM = 119.9\,pc\,cm^{-1}$.

\begin{table*}
\centering
\normalsize
\begin{minipage}{\linewidth}
\caption{\normalsize{Green Bank Telescope Summary of Observations}}
\centering
\begin{tabular}{|l|l|l|l|l|l|l|}
\hline
Obs No. & Sub-int No.& \pbox{40cm}{Obs. Start\\ Date} & \pbox{40cm}{Obs. Start \\ MJD} & \pbox{6cm}{Integration \\ Time (Mins.)} & \pbox{8 cm}{Orbital Phase \\ Coverage}\\
 \hline
\hline
\multicolumn{6}{|c|}{S-band observations}\\
\hline
1 & 1 & 2014-08-09 & 56878.149456 & 16.1 & 0.97-0.10\\
1 & 2 & 2014-08-09 & 56878.160641 & 16.1 & 0.10-0.24\\
1 & 3 & 2014-08-09 & 56878.171826 & 16.1 & 0.24-0.37\\
1 & 4 & 2014-08-09 & 56878.183011 &  11.6 & 0.37-0.46\\
\hline
2 & 5 & 2014-08-22 & 56891.022211 & 16.1 & 0.39-0.53\\
2 & 6 & 2014-08-22 & 56891.033396 & 13.8 & 0.53-0.65\\

\hline
\end{tabular}
\label{table:obssum}
\end{minipage}
\vspace{0.5 in}
\end{table*}

\section{Radio Data Analysis}
\label{section:DA}

We began the data analysis by sequentially combining groups of three
observational sub-integration of $322$\,s each (except last sub-ints
in both 9th and 22nd August observations which lasted for $55$ s and 
$187$\,s, respectively) . This resulted in four
independent raw data sets of ${\sim}16$\,min each, i.e. 13\% of
\saxj's binary orbital period of 2.01\,hr in each case.
Two raw datasets towards the end of $9$th and $22$nd August observations
were ${\sim}11$ and ${\sim}13$\,mins long, respectively.  
The total integration time was sub-divided in this way in order to
enable linear acceleration searches (see, \S \ref{section:BF}), and because of the
potential for eclipsing, which in analogy with the black widow systems
could last for at least 10\% of the orbit.  The observation start
times, duration, and the corresponding orbital phases of
\saxj\, are summarized in Table~\ref{table:obssum}.

Initial data preparation and periodicity searching was realized using
{\tt PRESTO}, a comprehensive pulsar processing software developed by
Scott Ransom \citep[for details see,][]{SCOTT:2001,REM:2002,
  RCE:2003}.  Radio frequency interference (RFI) was excised using an
RFI mask generated with {\tt rfifind}.  Given that the DM towards \saxj\, is unknown, we used {\tt
  prepsubband} to generate RFI-masked, barycentered, and de-dispersed
time series over trial DMs ranging from 0--1000$\rm\,pc\,cm^{-3}$
(using a DM step size of $0.1$\,pc\,cm$^{-3}$ up to a DM of
$500$\,pc\,cm$^{-3}$ and a step size of $0.3$ from 500--1000$\rm\,pc\,cm^{-3}$, resulting in $6671$ time series in total).  For
each time series we created a corresponding Fourier power spectrum
using {\tt realfft}.  The residual intra-channel DM smearing was
41--81$\,\mu$s (i.e. 1.6-3.2\% of the pulse period) for DMs of $100 -
200$\,pc\,cm$^{-3}$, which corresponds to an approximate distance
range of $3$--$6$\,kpc in the NE2001 model.

As described below, we searched the dedispersed time series for
pulsations using both a blind Fourier-based periodicity search {\it
  and} by directly folding the data using an X-ray-derived rotational
and orbital ephemeris.

\subsection{Blind Fourier-based periodicity search}
\label{section:BF}
We first performed a blind periodicity search in the event that the
X-ray derived ephemeris was inaccurate and to check the possibility for
a serendipitous, and unrelated radio pulsar along the line of sight.

The apparent rotational period of binary pulsars is Doppler shifted by
their binary motion. This results in spreading of spectral power over
multiple Fourier bins as $z = aT^2/cP$, where $z$ is the number of
Fourier bins drifted, $T$ is the integration length, $c$ is the speed
of light, and $P$ is the spin period.  For \saxj\,, the
maximum orbital acceleration is $a\approx\,14$\,m/s (companion mass $M_{\rm c}{\sim}0.05-0.08\,\msun$), corresponding to a drift of $z = 18$ bins in 16-min
observations.  As demonstrated by \citet{RGH:2001, REM:2002}, such a
signal can be successfully recovered by searching over multiple linear
frequency derivatives. We employed this technique of Fourier-based
acceleration searches, using {\tt accelsearch} and searched $z_{max} =
100$ for all the $6671$ Fourier power spectra (\S\ref{section:DA}) in
each $16$-min sub-integration.

We then identified the best candidates from the above acceleration
searches using the \textit{ACCEL\_sift} subroutine of {\tt PRESTO},
which groups candidates found at different trial DMs.
\textit{ACCEL\_sift} did not identify any candidates with a rotational
period close to that of \saxj\,.  Nonetheless, in case
there was a serendipitous pulsar along the same line of sight, for
each of the \textit{ACCEL\_sift} candidates, we folded the
corresponding de-dispersed time series and selected those folds
showing a reduced-$\chi^2 > 2$ (this is used as a proxy for
signal-to-noise) to also fold the raw data.  We then inspected the candidates
by eye and used parameters such as signal to noise ratio, measured DM, 
pulse profile and, converged period and period derivative solution from 
\textit{prepfold} output plot to make an informed selection. This inspection 
did not reveal any convincing pulsar candidate from the blind search.
 
\subsection{Direct Folding Search with X-ray-derived Ephemeris}

With {\it a priori} knowledge of the spin and orbital parameters, it
is possible to perform a deeper search for radio pulsations compared
to the blind search discussed above.  Previous coherent timing
analysis of \saxj\,, enabled by its X-ray pulsations during
outbursts, provides such an ephemeris.

However, the short-orbital-period black widow and redback millisecond
pulsar binaries are known to show non-deterministic orbital variations
(see, \citealt{pat12,bre13, AKH:2013}) and such variations should
also be expected in the case of \saxj, meaning that any previously
derived ephemeris may not extrapolate well to future observations.
X-ray pulsation searches in the redback transitional millisecond
pulsar PSR~J$1023+0038$ (e.g., \citealt{ABP:2015,jao16}) have
established that one can successfully account for such
non-deterministic orbital variations by searching over a small
deviation in the time of ascending node ($T_{\rm asc}$).  Therefore,
when folding the GBT radio data with X-ray derived ephemerides, we
searched both over DM and a $\Delta T_{\rm asc}$ value compared to the
fiducial ephemeris value.

Given the integration times of $1$ and $0.5$\,hr respectively during
the first and the second observation epochs, we could ensure
significant orbital coverage of \saxj\,'s $\sim$2\,hr orbit (see
Table~\ref{table:obssum}). We used two known orbital ephemerides: the
one obtained from coherent timing analysis up to 2011 \citep{pat12}
and the one obtained by also including the 2015 outburst (\S 4.2).  In
addition, we varied $T_{\rm asc}$ over a range of $\pm30$\,s in steps
of $0.1$\,s, resulting in $2\times600$ trial ephemerides per DM trial.

Each of the $6671$ dedispersed time series for every $16$-min
sub-integration were then folded using {\tt prepfold} and these
$2\times600$ ephemerides.  Moreover, the folding operation was
conducted in two additional ways: by allowing {\tt prepfold} to
optimize the S/N in a narrow range of spin period and spin period
derivative around the nominal ephemeris prediction and only
allowing an optimization in spin period derivative.  Hence, at the end
of these ephemeris-based searches we obtained
$2\times2\times6671\times600 = 16,010,400$ folded profiles. We
filtered the profiles by creating histograms of the S/N of the folds
in each 16-min sub-integration and choosing only candidates above a
certain threshold to inspect by eye.  
We found no candidate profiles with sufficient S/N that clearly peaked
in both trial DM and $\Delta T_{\rm asc}$.

\section{Results}

\subsection{X-Ray Lightcurve}

\begin{figure*}[t]
  \begin{center}
    \rotatebox{0}{\includegraphics[width=0.95\textwidth]{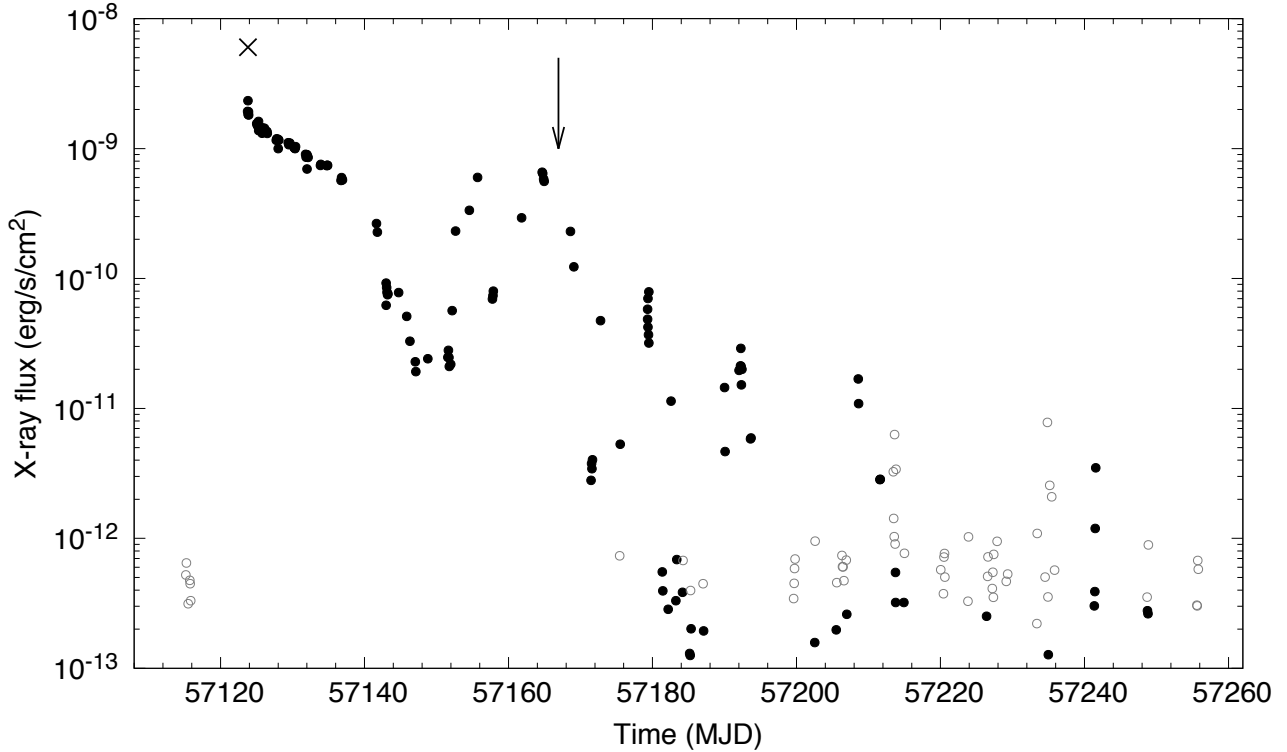}}
  \end{center}
  \caption{X-ray lightcurve (0.3--10 keV) of \saxj\, obtained with the
    \textit{Swift}/XRT telescope. The cross marks the occurrence of a
    Type I X-ray burst, whereas open circles are
    non-detection. The arrow identifies the 
    time of the \textit{Chandra} observation.}\label{fig:lc}
\end{figure*}

\begin{figure}[t]
  \begin{center}
    \rotatebox{-90}{\includegraphics[width=0.7\columnwidth]{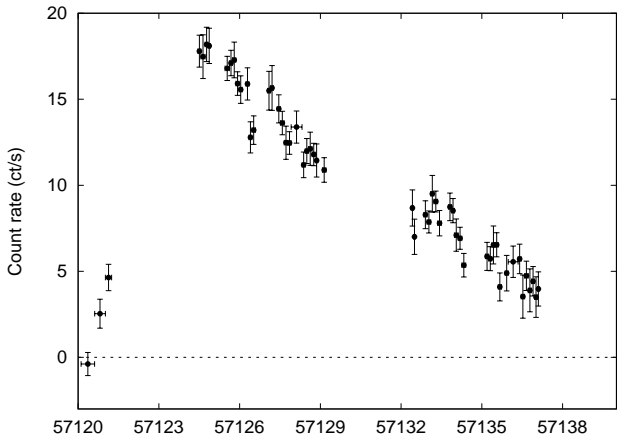}}
  \end{center}
  \caption{\textit{INTEGRAL}/IBIS lightcurve of the 2015 outburst in the 20--100 keV energy band. The onset of the outburst is detected around MJD~57121. The data points have typical integration times of 1800 -- 3600 s. The horizontal error bars (larger than the symbols only for a few points) define the time interval over which the data has been integrated. }\label{fig:integral}
\end{figure}

\saxj\, was detected in outburst with {\it Swift}/BAT on April 9th,
2015 (MJD 57121,~\citealt{san15}).  During the closest previous
\textit{Swift}/XRT observation, which occurred on April 3th (MJD
57195), \saxj\, was still in quiescence~\citep{cam15}.
The 0.3--10 keV X-ray lightcurve of \saxj\, (see Figure~\ref{fig:lc})
shows the very typical evolution that was also observed in the other
outbursts. The outburst has started after approximately 3.5 years
since the previous one, in line with the typical recurrence time of
3--4 years.

The {\it Swift}/XRT started monitoring \saxj\, after a Type I
 X-ray burst on April 11th (MJD 57123). The source
showed the same evolution seen in previous outbursts, with an observed
0.3--10 keV peak flux of ${\approx}3\times10^{-9}\ergs\,cm^{-2}$ that
(assuming a distance of 3.5 kpc; \citealt{gal06}) corresponds to a
luminosity of $4\times10^{36}\ergs$.  The outburst showed an initial
near exponential decay (slow decay) lasting about 15 days. It then
transitioned into a faster linear decay for about 5 days when the
source reached a luminosity of ${\approx}10^{35}\ergs$, before
entering a prolonged outburst reflaring tail that lasted another
$\sim100$ days.

 During the outburst reflaring tail, typical of all previous outbursts
 \citep{van00,wij01,wij03,cam08,pat09c,pat16}, the luminosity of the source
 oscillates between very faint states close to $2\times10^{32}\ergs$
 and relatively brighter ones of ${\sim}10^{36}\ergs$.  Two bright
 reflares were seen in 2015 on May 13th (MJD 57155) and May 18th (MJD
 57164) after which several progressively weaker reflares followed on a
 cadence of five to ten days. The {\it Chandra} observation we report 
 in this work took place during the second bright reflare.

 The power spectra of the \chandra\, data show no relevant feature at any 
 frequency. We exclude the presence of a 1~Hz modulation (similar to 
 that observed in several previous outbursts) with rms amplitude larger 
 than 10\% at the 95\% confidence level. This upper limit is derived
 by looking at the power in the 0.05--10 Hz range as done for example
 in \citet{pat09c}.

\subsection{X-Ray Pulsations}\label{sec:pulse}

The X-ray pulsations are very clearly detected in each data segment at
the 4-8$\sigma$ level, where we define the significance as the ratio
between the pulse amplitude and its statistical error. The sinusoidal
fractional amplitude of the pulsations is, on average, around 2\%
(sinusoidal amplitude or semi-amplitude) and does not show any
significant variation during the duration of the observations.  The
pulse time of arrivals are fitted with a constant pulse frequency plus
a Keplerian circular orbit and the statistical errors on the fitted
parameters are obtained with standard $\chi^{2}$ minimization
techniques.

Since we do not see any significant timing noise in the data (at the
timescales of the observations) and the variance of the pulse ToAs is
compatible with that expected from measurement errors alone, we take
our statistical errors as a good representation of the true
\textit{statistical} ones. In previous work (e.g., \citealt{har08,
  pat12}) it was shown that when observing the pulsations of \saxj\, a
strong timing noise is always observed on timescales of the order of
hours to days. Part of this noise is correlated to X-ray flux
variations and introduce systematic errors on the determination of the
spin frequency of the order of $10^{-8}$--$10^{-7}\rm\,Hz$.  These
systematic errors are particularly pronounced during the reflares when
strong pulse shape variability is observed~\citep{har08,har09}.  The
magnitude of such systematic errors can be estimated by looking at
long data stretches that are longer than the typical timing noise
timescales. However, since the pulsations available in our analysis
refer only to a short data span ($\approx 30$-ks) we cannot determine
the size of the systematic errors in our analysis. Indeed,
\citet{har08,har09} and \citet{pat12} estimated the average systematic
error on the pulse frequency over the entire baseline of the
observations, which lasted for weeks/months. Here, instead, the much
shorter data span implies that the systematic effect of timing noise
can be substantially larger than average.  For example, by looking at
Figure~1 in \citet{har09} we see that on timescales of few hours the
timing noise can induce pulse phase shifts of the order of 0.1--0.3
cycles. For this work such a phase shift would translate in systematic
errors on the determination of the pulse frequency of up to a few
$10^{-6}$Hz.

Even if our statistical errors on the pulse frequency are large
($\sim10^{-6}\rm\,Hz$) we cannot neglect the effect of systematic
errors, although we can only use a rough estimate of its magnitude by
looking at the behavior of the pulsations recorded during previous
outbursts.  The orbital and pulse frequency solution is reported in
Table~\ref{tab:1}.
\begin{table}
\centering
\caption{\saxj\ Timing Solution for the 2015 Outburst }
\begin{tabular}{llll}
\hline
\hline
Parameter & Value & Stat. Error & Syst. Error\\
\hline
$\nu$ [Hz] & 400.9752067 & $1.1\times10^{-6}$ & ${\sim}10^{-6}$\\
$T_{\rm asc}$ [MJD] & 57167.025002 & $7\times10^{-6}$ & \\
$e$ &  $<0.003$ &  (95\% c.l.) & \\
$a_1\,\rm\,sin\it\,i^{*}$ (lt-ms) & 62.812 & $2\times10^{-3}$& \\
$P_{\rm\,b}^{*}$ (s) & 7249.156980 & $4\times10^{-6}$&\\
Eccentricity $e^{*}$ (95\% c.l.)& $<1.2\times10^{-4}$ & &\\
Epoch (MJD) &  52499.9602472 & &\\
\hline
\end{tabular}\label{tab:1}
\begin{tablenotes}
\item $^*$ these values are taken from \citet{har09} and are kept fixed during the fit.
\end{tablenotes}

\end{table}

\begin{figure*}[t]
  \begin{center}
    \rotatebox{-90}{\includegraphics[width=1.0\columnwidth]{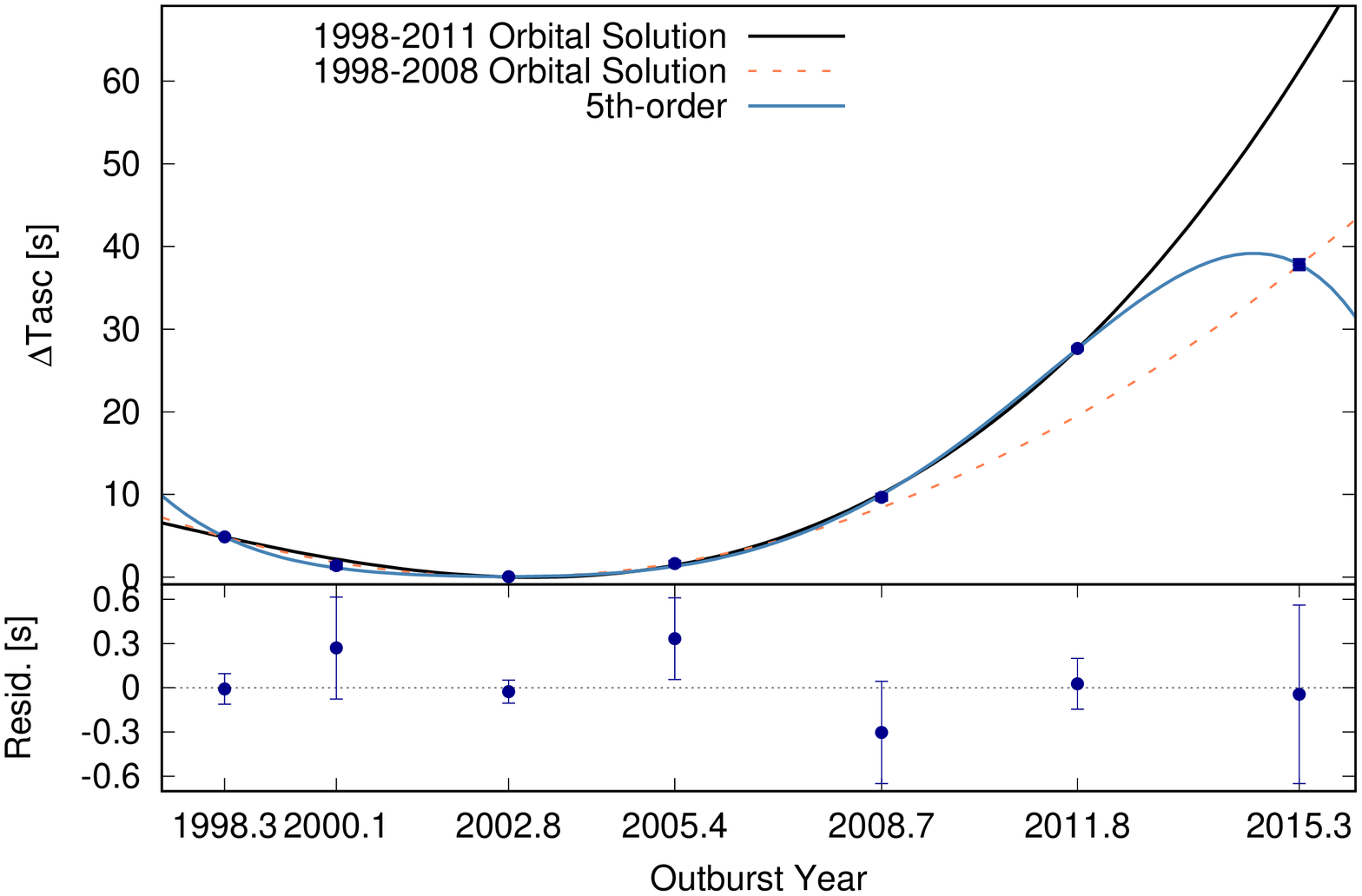}}
  \end{center}
  \caption{Orbital evolution of \saxj\, over 17 years. The
    $\Delta\,T_{\rm asc}$ cannot be fitted with a cubic (solid black line) or a
    quadratic (dotted orange line) polynomial. A 5-th order polynomial (solid blue line) is
    necessary to obtain a statistically acceptable fit, which suggests
    a complex behavior of the orbit. The panel in the bottom shows the residuals with respect to the 5-th order polynomial fit.}\label{fig:orb}
\end{figure*}

\subsection{Orbital Solution}\label{sec:orbsol}

To determine the orbital evolution we follow the procedure already used in
\citet{pat12, har09, har08} and fit the time of passage through the
ascending node (which is equivalent to orbital phase zero)
together with the measurements of the previous
outbursts.  The reference point $T_{\rm asc,ref}$ is taken from
Table 1 of \citet{har09}, and we use
the quantity $\Delta\,T_{\rm asc}=T_{\rm asc,i}-\left(T_{\rm asc,ref}+N\,P_{b}\right)$, where
$T_{\rm asc,i}$ refers to the $i$--th outburst and $N$ is the closest integer to
$\left(T_{\rm asc,i}-T_{\rm asc,ref}\right)/P_b$. The reference orbital period
$P_{b}$ can also be found in Table 1 of \citet{har09}.

Up to the 2008 outburst, the $\Delta\,T_{\rm asc}$ evolution showed 
a trend that was compatible with a quadratic polynomial 
representing an orbital expansion at a constant rate~\citep{har08,dis08}. 
Indeed, the time of passage through the ascending node can be expressed as
a polynomial expansion: 
\begin{equation}\label{eq:tasc}
T_{\rm asc}(N) = T_{\rm asc,ref}+P_bN+\frac{1}{2}P_b\dot{P}_bN^2+...
\end{equation}
By adding the 2011 outburst data points, \citealt{pat12} showed that a
quadratic polynomial was insufficient to describe the observed
behavior of the $T_{\rm asc}$ variations, which were instead successfully
described by a cubic polynomial. The physical interpretation given was
that, on the observed baseline of 13 years, the orbit was expanding at
an accelerated rate.

We now add the 2015 outburst data (see Figure~\ref{fig:orb}) and we
first try to fit the $T_{\rm asc}$ data points with a quadratic
polynomial, which corresponds to the solution found in \citet{har08,
  har09, dis08}. The fit is statistically poor with a $\chi^2$ of 492
for 4 degrees of freedom (dof).  A cubic polynomial is also a poor
description of the data with $\chi^2/\rm\,dof=314/3$. To obtain a p-value
above the canonical 5\% threshold we need to fit the data with a
fifth-order polynomial ($\chi^2/\rm\,dof=3.4/1$, p-value $6\%$), which
suggests that either the observed variability is governed by a
stochastic process or, if a periodicity is present, it must be
significantly longer than the observational baseline\footnote{As a
  cautionary test we also try to remove the 2011 point (assuming it is
  an outlier, even if there is no evidence or reason to believe that
  this is the case) and fit the data again with a quadratic
  polynomial. The data give also a poor fit with $\chi^2=11.1$ for 3
  degrees of freedom (and p-value $<1\%$)}.  We stress that the
concavity of the 5th order polynomial curve changes sign around 2011,
which implies that the orbit has started to shrink after that time.

Next, we tried to fit the data with a sinusoid that could represent
the effect of a slightly eccentric orbit with periastron advance. From
our previous work \citep{pat12} we already know that a sinusoid is a
statistically poor fit to the data. Indeed we find a formally bad fit
with a $\chi^2=83.9$ for 2 dof. Furthermore the fit requires an
eccentricity of about 0.004 which is much larger than the best upper
limits available on \saxj\, ($e<1.2\times10^{-4}$, \citealt{har09}).
Finally we attempted to fit the data with a Keplerian orbital delay
curve that could represent the effect of orbital motion caused by a
third body in an eccentric orbit. We find this can fit the data (with
$\chi^2=1.1$ for 1 dof) if the third body has a mass of about 8
Jupiter masses and is in a relatively wide ($\approx 5$ AU) orbit with
an eccentricity of about 0.7 and an orbital period of about 17.4
years. We can test this scenario by looking at the pulse frequency
derivative of \saxj\, since the pulsar would be accelerated along the
orbit. Given the fitted orbital parameters, the orbital velocity would
be $v_{orb}\approx 50\rm\,m\,s^{-1}$ with variations along the orbit
due to the large eccentricity. To get a first order orbital
acceleration we use $e=0$ and we get
$a_{orb}\approx6\times10^{-7}\rm\,m\,s^{-2}$. This would imply a
pulse frequency derivative of~\citep{jos97}:
\begin{equation}
\dot{\nu}_p = \nu_s\frac{\bm{a}_{orb}\cdot\bm{n}}{c}
\end{equation}
where $\bm{n}$ is a unit vector along the line of sight and $\nu_s=401$ Hz
is the spin frequency of \saxj\,. This gives $\dot{\nu}_p=8\times10^{-13}\rm\,cos\it\,\theta\rm$,where $\theta$ is the angle between the acceleration and line of sight vectors. Since we know from previous observations that \saxj\, is spinning down at a relatively constant rate of $\dot{\nu}_s\approx 10^{-15}\rm\,Hz\,s^{-1}$ we can confidently exclude this scenario.

\subsection{Radio Pulse Search}

Exhaustive searches using both a blind Fourier-based periodicity
search, and folding with a range of perturbed ephemerides, failed to
find radio pulsations from \saxj\, for any trial DM or
$\Delta T_{\rm asc}$ \S\ref{section:DA}.

In the absence of detectable radio pulsations, we can place a
stringent upper limit on pulsed radio emission from \saxj\,,
with the notable caveat that an active radio pulsar could in principle
be enshrouded by intra-binary material for a large fraction of the
time (e.g. \citealt{ABP:2015}).  In analogy with the black widow systems,
however, it is reasonable to assume that \saxj\, would only
be eclipsed for ${\sim}10$\% of its orbit at $2$\,GHz observing
frequency.

To set an upper limit on the flux density, we use the modified
radiometer equation \citep[see][]{DTW:1985, Bhat:1998, LK:2012}:

\begin{equation}
S_{min} = \frac{\left(\frac{S}{N}\right)~\beta~T_{sys}}{G~\sqrt{n_p~t_{obs}~\bigtriangleup f}}~\sqrt{\frac{W}{P-W}}
\end{equation}

We used the S-band receiver/frontend (Rcvr2\_3) at GBT. For this
receiver the system noise temperature $T_{sys}$ is $22$\,K and the
gain of the telescope $G$ is $1.9$\,K\,Jy$^{-1}$. Here,
$\bigtriangleup f$ is the $800$\,MHz bandwidth, the correction factor
$\beta$ is assumed ideal and close to $1$, the number of polarisations
$n_p$ is two and finally the integration time $t_{obs}$ corresponds to
$16$\,min\footnote{While in principle we could quote a 2x deeper limit
  by coherently folding the full 1-hr 9 August 2014 data set, we
  choose not to do so because some fraction of this integration is
  during orbital phases in which any radio pulsar is likely to be
  eclipsed.  We therefore prefer to set a more conservative flux
  density limit using the 16-min sub-integration, which together span
  a wide range of orbital phases.}.  With the assumption that the
pulse duty cycle is $\sim10\%$ and signal to noise ratio
$\left(\frac{S}{N}\right)_{min}$ for candidate identification by eye
is $8$, we obtain a maximum flux density of $30$\,$\mu$Jy at $2$\,GHz
(equivalently $\sim 50$\,$\mu$Jy at $1.4$\,GHz, for an assumed
spectral index of $\alpha = -1.4$).

This limit can be used as an important input for future radio
searches, and a point of comparison in the event that
\saxj\, becomes a detectable radio pulsar in the future.
We note that of the 106 Galactic field pulsars (outside of globular
clusters) in the ATNF catalog (accessible at http://www.atnf.csiro.au/research/pulsar/psrcat/;  See also \citealt{man05}) with quoted flux densities at
$1.4$\,GHz, and with spin period $< 10$\,ms, less than 10\% have
comparably low flux density to the upper limit we set on
\saxj\,.  These low-flux-density millisecond pulsars are
predominantly recent discoveries from the PALFA pulsar survey with
Arecibo \citep{SKL:2015, LBH:2015}.

\section{Discussion}\label{sec:discussion}

We have conducted the deepest radio pulse search for \saxj\, during
its quiescent phase in August 2014. No radio pulsations have been
detected, setting the strongest possible upper limit (30 $\rm\mu$J at
2 GHz) on the presence of radio pulsations that exist (to date) for
any AMXP. The presence of a radio pulsar turning on during quiescence
cannot be excluded with the present upper limits, but if a radio
pulsar signal is present it has to be quite weak at high radio
frequencies (2 GHz), substantially scattered by the intervening
interstellar medium, or perpetually eclipsed to be still compatible
with the current constraints. The beam width of millisecond radio
pulsars is very large (typical values of ${\sim}100^{\circ}$,
e.g.~\citealt{lor08}) so that missing the pulsar because of beaming,
although possible, is unlikely. The strongest evidence for a large
beaming angle comes from X-ray observations of globular clusters,
where very few unidentified X-ray sources have spectral properties
compatible with unknown millisecond pulsars whose radio beam is not
pointing towards Earth~\citep{hei05}. Even if \saxj\, is an active
radio pulsar in quiescence, there is still a good chance that eclipses
might appear for 10--50\% of the orbit due to free-free absorption by
intra-binary material, a common occurrence in black widow and redback
pulsars~\citep{nic00,rob13}.  To avoid this problem we have observed
at a high-enough radio frequency that a long eclipse duration is
unlikely.  We have also observed at a wide range of orbital phases,
when the neutron star is not behind its companion.

After 17 years of X-ray monitoring, the orbital period evolution of
\saxj\, shows a non-predictable behavior. The statistical fit to the
data show that neither a parabolic nor a cubic polynomial can describe
the data correctly. We found an ambiguity in the interpretation of the
long term trend of $T_{\rm asc}$, since the observations can be
explained in two ways. Either the orbit is expanding throughout the
17-years long observational window, with some fluctuations around the
mean $\dot{P}_b$, or the orbit has expanded until ${\sim}2011$
followed by a shrinkage (i.e., the fifth-order polynomial curve changes
concavity).  This is not a surprising behavior since many binary
systems have shown a similar orbital evolution. However, identifying
the precise short-term mechanism responsible for such orbital
evolution is a relatively difficult task.

In the following we will proceed by first discussing some fundamental
properties of binary evolution, then we will compare \saxj\, to other
known binaries that show anomalous orbital evolution and finally we
will review possible mechanisms to explain such an anomaly.

\subsection{Binary Evolution Timescales}\label{sec:fund}

Looking at the binary evolution, it is useful to define a
timescale $\tau_{\rm ev} = \frac{P_b}{\dot{P}_{b}}$ that can be
compared to the expected evolutionary timescales from theoretical
models. Differentiating the third Kepler law and assuming that all
mass lost by the companion is accreted by the primary,
one obtains the well known equation (see e.g., \citealt{fkr02}):
\begin{equation}\label{eq:orb}
\frac{\dot{a}}{a} = \frac{2\dot{J}}{J}+\frac{-2\dot{M}_{\rm c}}{M_{\rm c}}\left(1-q\right)
\end{equation}
where $q=M_{\rm c}/M_{\rm NS}$ is the mass ratio between the companion ($M_{\rm c}$)
and neutron star mass ($M_{\rm NS}$), $a$ is the orbital separation, $J$
is the angular momentum and the dot refers to the first time
derivative.
In general the angular momentum loss of the binary ($\dot{J}$) can be
decomposed in four terms (see e.g., \citealt{tau06}):
\begin{equation}\label{jdot}
\frac{\dot{J}}{J} = \frac{\dot{J}_{\rm gw}}{J}+\frac{\dot{J}_{\rm mb}}{J}+\frac{\dot{J}_{\rm ml}}{J}+\frac{\dot{J}_{\rm soc}}{J}
  \end{equation}
where the subscripts gw, mb, ml and soc refer to gravitational wave
emission, magnetic braking, mass loss and spin-orbit coupling,
respectively. When the binary is relatively compact, (orbital period of
less than $\sim1$ day), the evolution of the system is believed to be
driven by angular momentum loss (encoded in the $\dot{J}$ term in the
expression above; see e.g.~\citealt{fkr02}) rather than the nuclear
evolution of the donor star. In ultra-compact ($P_{b}<80$ min) and
compact binaries (80 min$<P_{b}\lesssim\,3.5$ hr) the angular momentum
loss is believed to be mainly due to emission of gravitational waves
($\dot{J}_{\rm gw}$) which becomes very efficient at short orbital separations (generally
when $P_{\rm b}\lesssim 3$hr, \citealt{vands11}). If there is no mass
loss from the system, then the loss of angular momentum via
gravitational waves drives the mass transfer and the orbital period
changes according to the following
expression~\citep{rap87,ver93,dis08}:
\begin{eqnarray}\label{eq:gw}
  \dot{P}_b = -1.4\times10^{-14}M_{\rm NS}&M_{\rm c}&M^{-1/3}P_{\rm\,b,hr}^{-5/3}\times\nonumber\\
  &\times&(\zeta-1/3)/(\zeta+5/3-2q)
\end{eqnarray}
where all masses are expressed in solar units, $P_{\rm b,hr}$ is the
orbital period in hours and $\zeta$ is the \textit{effective}
mass-radius index of the donor star ($R_{\rm c}\propto\,M_{\rm
  c}^{\zeta}$; see for example \citealt{vant98}).

When the orbital period of the binary is wider, in the range of
$3.5\rm\,hr\lesssim\,P_{b}\lesssim 0.5$--$1$ day, the dominant
mechanism driving the binary evolution is thought to be angular
momentum loss via magnetic braking ($\dot{J}_{\rm mb}$).  There is
currently considerable uncertainty about the details of magnetic
braking because its efficiency depends on a number of poorly
understood (and difficult to measure) stellar parameters (see e.g.,
\citealt{kni11} for a discussion). It also remains rather speculative
whether the single-star braking laws can be extended, unaltered, in
binary systems. Given these uncertainties the magnetic braking
timescale can vary by up to an order of magnitude and indeed different
recipes have been given in the literature
(\citealt{sku72,rap83,ste95}; see also \citealt{tau06} and Appendix A
in \citealt{kni11} for a review of several magnetic braking models).
Nonetheless, moderately wide binaries where magnetic braking is
dominant, are thought to lose angular momentum on a shorter timescale
than those compact binaries where angular momentum loss is dominated
by gravitational wave emission. The orbital parameters of \saxj\,
imply that it should be a typical gravitational waves driven binary,
since magnetic braking is believed to turn off (or become less
efficient) once the donor becomes fully convective and/or
semi-degenerate~(\citealt{spr83}; see however, \citealt{wri16} for
recent results that suggest that a dynamo process might still occur
in fully convective stars).

It is interesting at this point to compare what is observed in compact
radio pulsar binaries (black widows and redbacks, see
e.g.~\citealt{rob13}) as well as other LMXBs and accreting white
dwarfs which have similar orbital parameters as \saxj\, and have a
measured $\dot{P}_b$.  The reason why these systems might be relevant
in this context is two-fold: other LMXBs might be behaving in a
similar way as \saxj\, since the same mechanisms might be at play,
whereas black widows and redbacks are non-accreting systems and
therefore Eq.~(\ref{eq:orb}) simplifies. In the last few years three
redback pulsars have transitioned to an accreting LMXB state
\citep{arc09,pap13a,bas14,roy15}. Therefore it is still possible that
more of these systems (if not all) could display the same behavior
and therefore the assumption that redbacks are always non-accreting
might be invalid. As yet, no black widow system has been observed to
transition from a rotation-powered to an accretion-powered state.

\subsection{Comparison with Other Interacting Binaries}

A large number of interacting binaries have a measured orbital period
evolution which is too fast to be explained with simple binary
evolution models involving only $\dot{J}_{\rm gw}$ and $\dot{J}_{\rm
  mb}$ and requires a number of additional effects. These systems
include binary pulsars, cataclysmic variables and LMXBs with neutron
star and black hole accretors. Among the binary pulsars, we discuss
below only the cases of the black widows and redbacks because all
binaries with a white-dwarf/neutron star companion are following the
predictions of general relativity with exquisite precision (e.g.,
\citealt{tay89, wei02}). We also exclude from the sample those radio
pulsars with a B-type star companion (not to be confused with the Be
X-ray binaries, which are accreting systems and a subset of high mass
X-ray binaries) since very different mechanisms involving the short
nuclear evolution timescale of the massive companion need to be
considered. This is also the reason why we do not include high mass
X-ray binaries in our sample.

\subsubsection{Black Widows and RedBacks}
In black widows and redbacks the companion star is being ablated by
the pulsar wind and high energy radiation, thus producing potential
mass loss. Observational evidence of this phenomenon comes from the
fact that radio pulsations are very often eclipsed by intra-binary
material that induces free-free absorption of the pulsed signal. The
orbital parameters of \saxj\, are compatible with those of a BW, and
its 0.05-0.08$\msun$ companion is also a semi-degenerate
star~\citep{bil01,del08, wan13}. The only difference between BWs and
\saxj\, is that in the latter system the companion is in Roche lobe
overflow whereas BWs are thought, at least in some cases, to be detached systems~\cite{bre13}.

In black widows, as well as in redback pulsars, the
\textit{short-term} effects on the orbital evolution do occur on
timescales which are generally orders of magnitude shorter than the
predicted (secular) ones from angular momentum loss due to
gravitational waves and/or magnetic braking.  The six BWs and RBs
which have a measured (and publicly available\footnote{For a complete
  list of binary pulsars we refer to the ATNF pulsar catalog
  http://www.atnf.csiro.au/people/pulsar/psrcat/}) $\dot{P}_b$ show
orbital evolution timescales from 100 Myr down to less than 1 Myr (see
Table~\ref{tab:3}). Five of them have negative orbital period
derivative (the orbit is shrinking), whereas only one (PSR B1957+20)
has a positive value (the orbit is expanding).  Similarly, in the case
of \saxj, $\tau_{\rm ev}= \frac{P_b}{\dot{P}_{b}}\approx 70$ Myr,
whereas from Eq.~(\ref{eq:gw}) one would have expected a timescale of
a few Gyr (varying slightly with the exact neutron star and companion
mass chosen) thus meaning that the $\dot{J}_{\rm gw}$ term is not the
dominant one. It is important to stress that these apparent orbital
period derivatives are epoch dependent, and change on year-long
timescales.

\begin{table}
\centering
\caption{Black Widows and Redbacks with measured $\dot{P}_b$}
\begin{tabular}{lllll}
\hline
\hline
Name (PSR) & Type & $P_b (hr)$ & $\dot{P}_b$ & Error\\
\hline
J1023+0038 & RB &  4.8 & $-7.3\times10^{-11}$ & $0.06\times10^{-11}$\\
J1227-4853 & RB &  6.9 & $-8.7\times10^{-10}$ &$0.1\times10^{-10}$ \\
J1723-2837 & RB & 14.8 & $-3.5\times10^{-9}$ &$0.12\times10^{-9}$ \\
J1731-1847 & BW &  7.5 & $-1.08\times10^{10}$ & $0.07\times10^{-10}$\\
J2051-0827 & BW &  2.4 & $-1.6\times10^{-11}$ & $0.08\times10^{-11}$\\
J1959+2048 & BW &  9.2 & $ 1.47\times10^{-11}$& $0.08\times10^{-11}$\\
\hline
\end{tabular}\label{tab:3}
\end{table} 

\subsubsection{Other Low Mass X-Ray Binaries}
In the few compact LMXBs where an orbital period derivative can be
measured, we see a roughly equally distributed sign (6 positive, 4
negative signs and three upper-limits observed so far).  The magnitude
of the orbital period derivative is always much larger than expected
from conservative binary evolution ($\dot{J}=0$) and/or from angular
momentum loss via gravitational waves and/or magnetic braking. In the
literature a number of different mechanisms have been suggested to
explain the large $\dot{P}_b$.  These include mass loss from the
companion~\citep{bur10, pon15}, enhanced magnetic
braking~\citep{gon14}, the presence of a third body~\citep{iar15},
spin-orbit coupling~\citep{wol09,pat12} and in some cases even
modified theories of gravity~\citep{yag12}. The group of LMXBs
appears to be the most heterogeneous among the different binaries that
we are considering here. Indeed this group comprises transient LMXBs
with black hole accretors (XTE J1118+480 and A0620--00), transient
LMXBs with neutron star accretors (EXO 0748--676, MXB 1658--298, SAX
J1748.9--2021, AX J1745.6--2901 and \saxj\,) and persistent neutron star
LMXBs (Her X--1, 2A 1822--37, XB 1916--053). Furthermore, some of these
LMXBs are accreting pulsars and have orbital periods determined via
timing of their pulsations (e.g., Her X--1, 2A 1822--37, \saxj\, SAX
J1748.9--2021) whereas others are eclipsing systems and their period
is determined via X-ray and/or optical photometry. No other orbital
period derivative has been measured so far for any other LMXB. In
Table~\ref{tab:2} we list the LMXBs sample with their orbital period
derivatives and the main proposed explanation given in the literature.

The binary EXO 0748--676 is an eclipsing binary with a 0.4$\msun$
donor that shows sudden variations in ${\dot{P}}_b$, which were
proposed to be due to spin-orbit coupling~\citep{wol02,wol09}.
\citet{wol02} analyzed two segments of X-ray data (1985--1990 and
1996--2000) and showed that the period had increases by about 8
ms. However, the period increase shows jitters and cannot be fit with
a constant ${\dot{P}}_b$. Further work by \citet{wol09} extended
the analysis until 2008 and observed a similar behavior. 

The eclipsing binary 2A 1822--37 is an accreting pulsar with a
${\sim}0.3-0.4\msun$ companion and it has a relatively steady increase
of the orbital period measured over a baseline of 30 years (see e.g.,
\citealt{bur10,iar11,cho16}). This system is an accretion-disk corona
system showing extended partial eclipses of the central X-ray
source. \citet{bur10} and \citet{iar11} have suggested that the binary
contains an Eddington limited accreting neutron star whose irradiation
of the donor is inducing severe mass loss that can explain the large
orbital period derivative observed. The positive sign of ${\dot{P}}_b$
is ascribed to the response of the radius of the donor to mass-loss,
with $R_c\propto\,M_c^\zeta$ and $n<1/3$. \citet{bur10} suggests that the
donor in 2A 1822--37 has a deep convective envelope with $\zeta=-1/3$ (see
e.g., \citealt{rap82}) thus justifying the positive $\dot{P}_b$.

The source AX J1745.6--2901 is an eclipsing binary with an accreting
neutron star and a negative ${\dot{P}}_b$~\citep{pon15}. The donor
mass is constrained to be $M_c\lesssim0.8\msun$. A strong mass loss is
also suggested in this case, but since the system is shrinking the
mass-radius index needs to be $n>1/3$.  The data on the $T_{\rm asc}$
collected over a baseline of about 30 years show significant scatter
of up to several tens of seconds.

XB 1916--053 is an ultracompact ($P_b\approx 50$min) persistent dipping
source monitored for over 37 years by X-ray satellites. \citet{hu08}
studied the first 24 years of data and found that a quadratic function
(i.e., a constant ${\dot{P}}_b$) was able to describe the data
correctly. However, \citet{iar15} found that, when considering the
entire 37 years of observations, a quadratic function was unable to
fit the orbital evolution and a model with a sinusoidal variation in
addition to the quadratic component was required. A third body with
mass of $\sim0.06\msun$ was invoked to explain the observations with
an orbital period of $\approx 26$ years.  It is instructive to notice
that a deviation from a quadratic function was not apparent in the
first 24 years of data, which suggests that a very long timescale
periodicity (or quasi-periodicity) might still be present even in
binaries where a constant ${\dot{P}}_b$ is observed over baselines of
a few decades.

The globular cluster source SAX J1748.9--2021 is instead an
intermittent AMXP \citep{alt08,gav07, pat09,san16} and it has been
observed in outburst five times.  Its companion star is likely to be a
${\sim}0.8\msun$ star close to the turnoff mass of the globular
cluster NGC 6440, although much smaller masses down to 0.1$\msun$
cannot be excluded~\citep{alt08}.  The orbital evolution has been
studied by looking at the orbital ephemeris calculated with coherent
timing in a way similar to what has been done in this
work. \citet{san16} describes the orbital evolution with a quadratic
function although the fit shows large deviations of the order of 100
seconds from the best fit function (which translated into a poor
$\chi^2$ of 78.4 for 1 dof).  These authors interpret the large
orbital expansion with a highly non-conservative mass loss scenario
where the binary is losing more than 97\% of the mass flowing through
the inner Lagrangian point $L_1$.

Her X--1 shows instead a steady decrease of the orbit whose
value is however compatible with both a conservative and a non-conservative
mass transfer scenario~\citep{sta09}.

Finally, the orbital evolution of the two transient black hole LMXBs
XTE J1118+480 and A0620--00~\citep{gon12,gon14} was measured with
radial velocity curves determined via optical spectroscopy for a
period of time of $\sim10$ and 20 years respectively. These
observations were all carried out during the extended periods of
quiescence of the binaries. The orbit shows a steady shrinkage
interpreted as due to enhanced magnetic braking.  The two binaries
have a companion mass of 0.2$\msun$ (XTE J1118+480; \citealt{gon14})
and 0.4$\msun$ (A0620--00; \citealt{can10,gon14}), respectively.

\subsubsection{Cataclysmic Variables}
It is well known that some cataclysmic variables show an anomalous
orbital period derivative as well, with some of them proposed to be
transferring mass at a higher rate than expected due to irradiation of
the companion~\citep{patt16, patt15, kni00}. Even some Algol type
binaries (i.e., a semi-detached system composed by a detached early
type main sequence star and a less massive sub-giant/giant star in
Roche lobe overflow) have been reported to evolve on a very short
timescale \citep{erd14}. In this case non-conservative mass transfer
scenario is expected to take place since the red giant will emit a significant 
wind. However, for a few of these systems (i.e., all the
converging ones) the required mass loss is larger than the highest
theoretical value for wind mass loss in giant stars. These
observations might suggest that short term effects have some influence
on the orbital evolution of accreting and non accreting neutron stars,
persistent and transient systems and white dwarf/black hole/main
sequence stellar accretors. It is worth noticing that no neutron star
+ white dwarf binary (both accreting and non-accreting) has been
observed (as yet) to evolve on anomalous timescales. The only exception
is the ultra-compact LMXB 4U 1822--30, which, however, is located in a
globular cluster and therefore its large $\dot{P}_{b}$ might simply be
due to the contamination induced by the gravitational potential well of the cluster
\citep{jai10}. This suggests that, if there is a common reason behind
this behavior for all type of binaries (which is of course not
necessarily true), it must be related to the type of companion (main
sequence or semi-degenerate star) rather than the type of accretor.

\subsubsection{Caveats}
\begin{figure*}[t]
  \begin{center}
    \rotatebox{-90}{\includegraphics[width=0.7\textwidth]{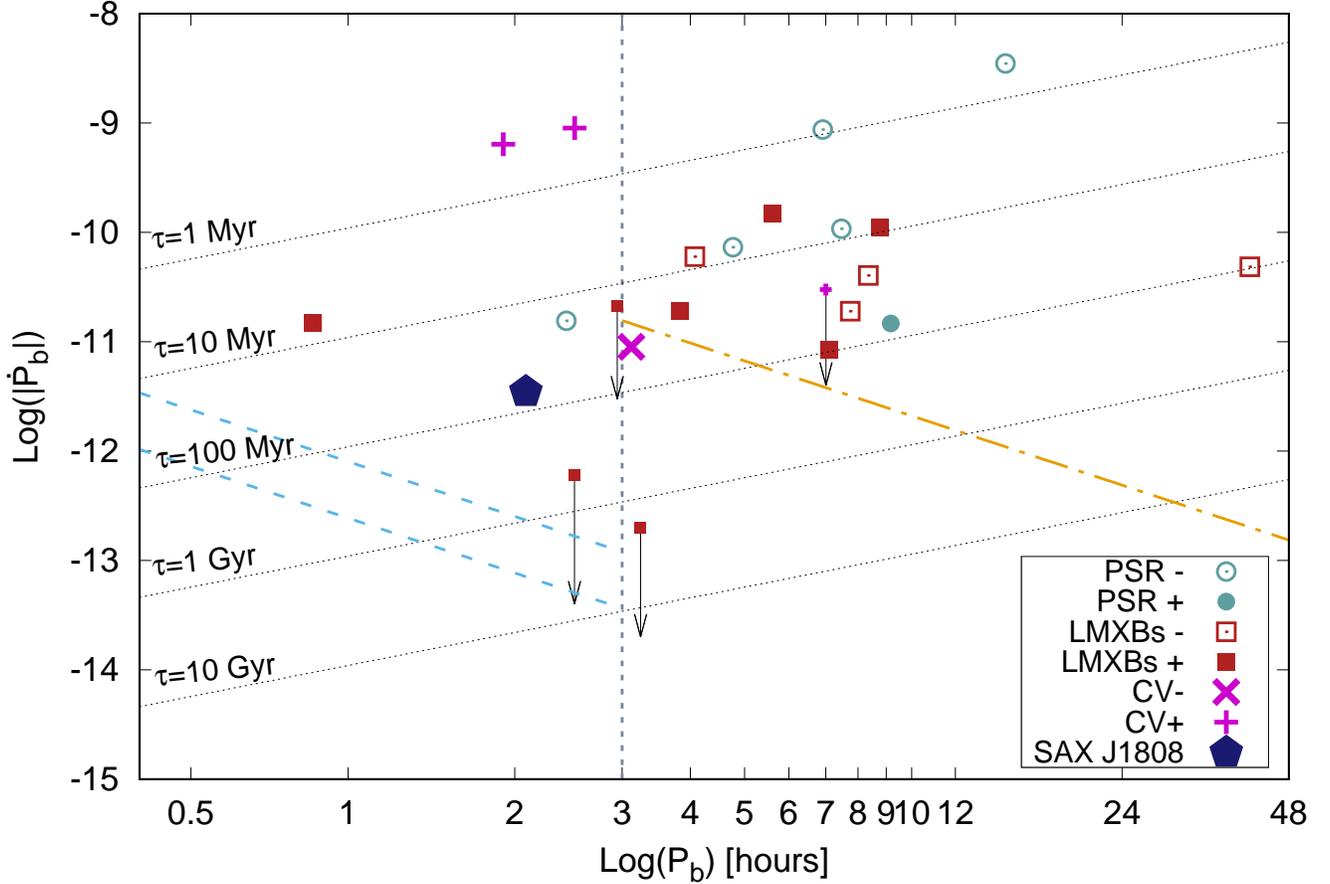}}
  \end{center}
  \caption{$|\dot{P}_b|$ vs. $P_b$ diagram for \saxj\, (blue
    pentagon), LMXBs (red squares), binary pulsars (BW and RBs, blue
    circles) and some CVs (purple crosses). The open symbols identify
    converging systems (negative $\dot{P}_b$) whereas filled symbols
    are diverging ones. Only T PyX and IM Nor are plotted for CVs
    (plus NN-Ser which is drawn with the same symbol used for
    CVs). The oblique black dotted lines identify evolutionary
    timescales $\tau=P_b/|\dot{P}_b|$. The dashed vertical lines
    roughly separates the binaries where the dominant angular momentum
    loss should be gravitational radiation (left) from those where
    magnetic braking is expected to dominate (right) assuming no mass
    loss or spin-orbit coupling is present in the binary.  The cyan
    oblique lines are theoretical values of $|\dot{P}_b|$ expected if
    gravitational wave emission is the main driver of orbital
    evolution. From top to bottom these lines are valid for a donor
    mass-radius relation $R_c\propto\,M_c$ and
    $R_c\propto\,M_c^{-1/3}$ respectively, assuming a donor mass of
    0.08$\msun$ and a neutron star mass of 1.4$\msun$. A value of
    $0.08\msun$ has been chosen as it represents the most likely donor
    mass in \saxj\,. The orange oblique lines is instead the orbital
    period derivative as a function of orbital period when the maximum
    possible magnetic braking effect is considered (see main text for
    a discussion). A mass-radius index $\zeta=1$ has been assumed for
    the magnetic braking case.}\label{fig:pbdot}
\end{figure*}

 A few cautionary words are necessary at this point on the CV, Algol
 type binaries and some LMXBs. For these type of systems, where the
 $\dot{P}_b$ is detected by looking at the eclipse times in optical
 data, some selection effects might be present.  This means that those
 systems with a $\dot{P}_b$ in line with the theoretical predictions
 might be more difficult to measure/detect and therefore the reported
 values are invariably skewed towards large/anomalous $\dot{P}_b$
 values.  Something similar applies also to most LMXBs, with the
 exception of those systems where the $\dot{P}_b$ is measured via
 pulsar timing, in which case the sensitivity of the timing technique
 potentially allows the detection of values orders of magnitude smaller than in
 the CVs and Algol binaries.  For the CVs there is ample literature on
 the topic and several different systems with a large $\dot{P}_b$ are
 reported.  In this paper we include only the T Pyxidis and IM Normae
 systems which are the two best studied cases and have the highest
 orbital period variation~\citep{patt15,patt16}. We also include
 NN-Serpentis~\citep{bri06} which is an eclipsing post common-envelope
 binary where no mass transfer is currently ongoing.  We summarize the
 information on the orbital period evolution of all the binaries
 discussed in this work in Figure~\ref{fig:pbdot}.  From the figure,
 it is clear that all sources with short orbital periods that should
 be losing angular momentum via gravitational wave emission, are
 evolving on timescales which are at least an order of magnitude
 shorter than expected. The binaries with wider orbits, where magnetic
 braking should dominate, show also shorter evolutionary timescales
 than predicted, although a larger scattering is observed and some
 sources are close to the theoretical predictions.

\begin{table*}
\begin{threeparttable}
  \centering
\caption{Binaries with Anomalous Orbital Period Derivatives}
\begin{tabular}{llllllll}
\hline
\hline
Name & $P_{b} [hr]$ & $\dot{P}_{b}$ & Transient & Companion Type & Sign & Proposed Model & References\\
\hline
Neutron Star LMXBs\\
\hline
EXO 0748--676$^{a}$	 & 3.8 & $1.9\times10^{-11}$  & Yes  & MS       & + & SOC           & \citet{wol09, wol02}\\
2A 1822--37	 & 5.5 & $1.51(8)\times10^{-10}$  & No   & MS       & + & Mass Loss           & \citet{bur10,iar11}\\ 			
SAX J1808.4--3658 & 2.0 & $3.5(2)\times10^{-12}$  & Yes  & SD       & + & SOC/Mass Loss & \citet{pat12,dis08}\\
MXB 1658--298     & 7.1 & $8.4(9)\times10^{-12}$  & Yes  & MS       & + &  Unknown             & \citet{pau10}\\
XB 1916--053      & 0.8 & $1.5(3)\times10^{-11}$  & No  & SD       & + &  Third Body         & \citet{iar15}\\
SAX J1748.9--2021 & 8.8 & $1.1(3)\times10^{-10}$  & Yes  & MS/Sub-G & + &  Mass-Loss       & \citet{san16}\\
AX J1745.6--2901  & 8.4 & -$4.03(32)\times10^{-11}$  & Yes  & MS/Sub-G       & -- & Mass Loss      & \citet{pon15}\\
Hercules X--1     & 40.8& -$4.85(13)\times10^{-11}$  & No   & MS       & -- & Several          &\citet{sta09}\\
\hline
Black Hole  LMXBs\\
\hline
XTE J1118+480 & 4.0 & -$6(1.8)\times10^{-11}$ &  Yes & MS & -- & Enhanced MB & \citet{gon12,gon14}\\ 
A0620--00      & 7.8 & -$1.9(3)\times10^{-11}$ &  Yes & MS & -- & Enhanced MB & \citet{gon14}\\
\hline
\end{tabular}\label{tab:2}
\begin{tablenotes}
\item MS = main sequence; sub-G = sub-giant;  SD = semi degenerate; MB= magnetic braking; SOC = Spin-Orbit Coupling;
\item $^{a}$ This source shows segments of data where a constant $P_b$ is required. The error on $\dot{P}_b$ is not given and confidence intervals are determined via Maximum Likelihood Method~\citep{wol09}.
\end{tablenotes}
\end{threeparttable}
\end{table*}

\subsection{Models}

Since a large number of scenarios are invoked in the literature to
explain the orbital evolution of different interacting binaries, it
appears legitimate to ask whether it is
still possible to find a common (and/or perhaps still unknown) mechanism
behind the observed behavior.  In the
following discussion we proceed by considering all models proposed in
the literature and try to apply each of them to the case of \saxj\,.
with the exception of the third body model, which has be already
excluded (see Section~\ref{sec:orbsol}.

\subsubsection{Mass Loss Model}\label{sec:ml}

If the companion star experiences severe mass loss, for example
because of ablation due to the irradiation from a pulsar wind or from
the X-rays originating close to the compact object, then the orbital
period of the binary changes dramatically (see
Section~\ref{sec:fund}).

In this case the orbital period derivative will depend on the amount
of wind lost from the companion:
\begin{equation}\label{ml}
\frac{\dot{P}_b}{P_b} = -2\frac{\dot{M}_{\rm c}}{M_{\rm c}}
\end{equation}
where $M_{c}$ is the companion mass (see e.g., \citealt{fkr02, pos14}). 
Applying this model to \saxj\, it is possible to explain the observed $\dot{P}_{b}$
if the donor is losing mass at a rate of about $10^{-9}\rm\,\msun\,yr^{-1}$~\citep{dis08}. \\

For the donor to lose a substantial amount of mass, there must be a
way to efficiently inject energy into the donor star. Whatever
mechanism is chosen, the amount of energy necessary to create such a
strong mass loss must be consistent with the total energy budget
available to the binary. In \saxj\, it has been proposed that the mass
loss is driven by a pulsar wind and high energy radiation impinging
onto the donor surface~\citep{dis08, bur09}.  For a circular binary
orbit, the total angular momentum is~\citep{fkr02}:
\begin{equation}\label{jorb}
J \propto M_{\rm NS}M_{\rm c} M^{-1/3}P_b^{1/3}      
\end{equation}
where $P_b$ is the orbital period and $M = M_{\rm NS} + M_{\rm c}$ the
total binary mass, and $M_{\rm c} < M_{\rm NS}$.  The orbital energy $E_{\rm orb}$
is

\begin{equation}\label{eorb}
-E_{\rm orb} \propto J/P_b                                         
\end{equation}

From Eq.~\ref{jorb}, an increase in $P_b$ requires $M_{\rm c}$ to decrease,
since for an isolated system $J$ and $M$ cannot increase.  And from
Eq.~\ref{eorb} we see that $-E_{\rm orb}$ must decrease, making the orbit less
tightly bound. In other words, an orbital period increase requires
energy injection from somewhere.
\citet{mar90} show that energy injection by the secondary star
is too slow for observed $P_b$ changes as this is governed by the star's
thermal timescale\footnote{This is of course what happens when a very
  low-mass star expands on mass transfer ($M_{\rm c} —> M_{\rm NS}$) with $P_b
  \propto 1/M_{\rm c}$. The star’s thermal energy expands it adiabatically.}.

The only energy source left is the spin energy of the NS.
This is:
\begin{equation}
E_{\rm spin} \sim k^2\,M_{\rm NS}\,v_K^2 \sim k^2\,GM_{\rm NS}^2/R_{\rm NS}
\end{equation}
where k is the radius of gyration and $R_{\rm NS}$ its physical radius, and $v_K
= (GM_{\rm NS}/R_{\rm NS})^{1/2}$ is the breakup spin velocity.  This is a huge
reservoir, since
\begin{equation}
E_{\rm spin}/(-E) \sim k^2\frac{M_{\rm NS}}{M_{\rm c}}\frac{a}{R} \sim 400 - 1000 
\end{equation}
with $k^2 \sim 0.4$. But of course not all of $E_{\rm spin}$ can be used
to drive mass loss from the system and so increase $P_b$.  We can estimate
the required minimum efficiency $\eta$ for spin energy conversion into
orbit energy for \saxj\,.

The total orbital binding energy of the binary is:
\begin{equation}
E_{\rm orb}=-\frac{GM_{\rm NS}M_{\rm c}}{2a}
\end{equation}
If we use the third
Kepler law, then we have that the orbital period is very well
determined ($P_b=2.01$ hr, from the X-ray pulse timing). The total
mass of the binary is ill constrained, mostly because of the unknown
neutron star mass. If we assume a range of total binary mass
from 1.4 up to 3$M_{\odot}$, then the variation in $a$ is of the order
of 20\% ($6-8\times10^{10}\rm\,cm$). Here we will assume
$a=6.4\times10^{10}\rm\,cm$ ($M_{\rm NS}=1.4\msun$, $M_{\rm c}=0.08\msun$).
The orbital energy is therefore:
\begin{equation}
E_{\rm orb}\sim3\times10^{47}\rm\,erg
\end{equation}
The semi-major axis of the binary changes according to the 3rd Kepler law:
\begin{equation}
\dot{a}=\frac{2a}{3P_b}\dot{P}_{b}
\end{equation}
The value $\dot{P}_b$ is measured from observations and its value is
${\sim}3.5\times10^{-12}$. Therefore
$\dot{a}=2\times10^{-5}\rm\,cm\,s^{-1}$.  The orbital energy variation
is (we assume $\dot{M}$ terms are negligible):
\begin{equation}
\dot{E}_{\rm orb}= \frac{GM_{\rm NS}M_{\rm c}\dot{a}}{2\,a^2}\sim9\times10^{31}\rm\,erg\,s^{-1}
\end{equation}
The total spin down power is:
\begin{equation}
\dot{E}_{\rm sd}=I\omega\dot\omega  
\end{equation}
where $\omega=2\pi\,\nu$, $\dot\omega=2\pi\dot\nu$ and $\nu$ and $\dot\nu$
are the spin frequency (401 Hz) and the spin down ($1.65(20)\times10^{-15}\rm\,Hz\,s^{-1}$)
observed in \saxj\,~\citep{pat12}.
Numerically, $\dot{E}_{\rm sd}\approx2.6\times10^{34}\ergs$.

We need an efficiency $\eta$ of at least the ratio between the two powers:
\begin{equation}
\eta=\frac{\dot{E}_{\rm orb}}{\dot{E}_{\rm sd}}\sim0.003
\end{equation}

For the parameters assumed above, the Roche lobe radius is $R_L=0.16\,R_{\odot}$~(see e.g.,
\citealt{egg83}) and the fraction of intercepted power is
$f=(R_L/2a)^2$ which is approximately $0.8\%$.  This value suggests
that the donor star must be extremely efficient in converting the
incident power into mass loss since
$\epsilon\sim\eta/0.8\approx 40\%$.  \citet{cam04} estimated that the
irradiating power\footnote{The authors used a distance of 2.5 kpc,
  whereas here we rescale the luminosity for a distance of
  $3.5\pm0.1$~kpc as determined by \citet{gal06}} required to explain
the bright optical counterpart of \saxj\, (observed with VLT data in
2002) amounted to $L_{\rm irr}=8^{+3}_{-1}\times10^{33}\ergs$ (see
also \citealt{hom01} for a similar estimate made with observations taken in
1999). Therefore the optical data require a conversion of a fraction
$\xi=L_{\rm irr}/\dot{E}_{\rm sd}\approx0.2$--$0.6$ of incident
  power into thermal radiation by the donor star (the range provided
takes into account the $1~\sigma$ error bars on the irradiation
luminosity and the possibility that the distance is 2.5~kpc rather
than 3.5~kpc).  Since from the observational constraint we have that
$\xi+\epsilon=0.5-1.0$ and $\xi+\epsilon$ is
bound to be equal to $1$ by the conservation of energy, this scenario is
energetically plausible for a range of parameters compatible with the
observations.  

\begin{figure}[t]
  \begin{center}
    \rotatebox{-90}{\includegraphics[width=0.35\textwidth]{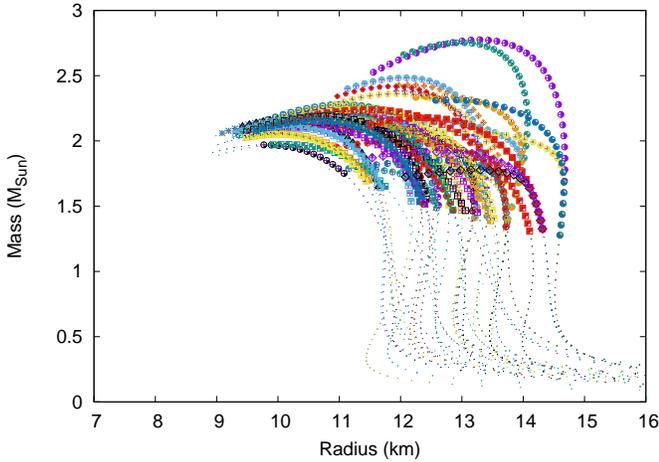}}
  \end{center}
  \caption{Constraints on the mass-radius relation of the neutron star in \saxj\, under the assumption that the binary is driven by mass loss. The different curves correspond to different equation of states of ultra-dense matter and are taken from \citet{for16}. The thick part of the curves marks the segments for which the moment of inertia of the neutron star is $>1.7\times10^{45}\rm\,g\,cm^{2}$ (see main text for a discussion). }\label{fig:eos}
\end{figure}

If we assume that the mass loss from the donor of \saxj\, is constant,
then one still needs to explain the $T_{\rm asc}$ of the 2011
outburst.  This data point deviated by approximately 7 seconds from
the predicted value that can be obtained in Eq.~(\ref{eq:tasc}) by using
a constant $\dot{P}_b=3.5\times10^{-12}\rm\,s\,s^{-1}$.  The 7-s
deviation can be explained if the mass loss rate has increased by
about $\sim70\%$ during the 2008-2011 period, which would require a
proportionally larger spin-down power than assumed above. Indeed the
relation between the mass loss and the spin-down power is linear (see e.g.
a discussion in \citealt{har09}):
\begin{equation}
  \dot{M}_{\rm c}=\dot{E}_{\rm sd}\left(\frac{1}{2a}\right)^2\,G\,M_{\rm c}\,R_{\rm c}
\end{equation}
Since we have seen that the donor needs to convert the incident
spin-down power into mass-loss with extraordinary efficiency, close to
40\% ($\epsilon\sim0.4$), then the spin-down power is larger than
initially estimated either because of a larger moment of inertia of
the neutron star or because the spin-down $\dot\omega$ is slightly
larger than observed. In the first case one would need
$I\gtrsim1.7\times10^{45}\rm\,g\,cm^2$ and this can be used in
principle to constrain the equation of state of ultra-dense matter
(under the assumption that mass-loss is the main mechanism responsible
for the binary evolution). In Figure~\ref{fig:eos} we plot an
illustrative example of the type of constraints that can be obtained
when using this method for a selection of equations of
state~\citep{for16}.

In conclusion, this mechanism is energetically feasible if:
\begin{itemize}
\item \saxj\, has a neutron star with a large moment of inertia
\item the incident pulsar power can be converted into mass loss with
  a $\approx40\%$ efficiency. 
\end{itemize}

When looking at the whole sample of pulsar binaries, the behavior of
other black-widow pulsars cannot be explained by this model since at
least two of them are \textit{shrinking} their orbit.  Therefore if
the mass-loss model is correct we need two different mechanisms to
explain the population of binary pulsars.  Furthermore, in a recent
work (Patruno 2016) we have studied the orbital evolution of
another AMXP, namely \igr\,.  This system can be considered a ``twin''
system of \saxj\, since its orbital and physical parameters are
extremely similar (see e.g., \citealt{pat12r}).  However, in that case
the orbit of the binary is varying at a very slow pace ($\tau>0.5$
Gyr), compatible with a conservative scenario where the binary
evolution is driven by gravitational wave emission. In that case
we constrained the
efficiency of the pulsar spin-down to mass-loss conversion to be
$\lesssim 5\%$. It is not clear therefore why \igr\, is unable to
convert spin-down power into mass-loss whereas \saxj\, is so
efficient. We stress that the observations of \textit{both} systems
suggest a donor irradiated by a pulsar wind/high energy radiation and
the donor mass is almost identical.

\subsubsection{Spin-Orbit Coupling}

An exchange of angular momentum between the stellar spin and the orbit
can generate variations of the orbital period.
This variations of the orbital angular momentum are encoded in the
term $\dot{J}_{\rm soc}$ in Eq.(\ref{jdot}).

The so-called
Applegate model was developed by~\citet{app92} and \citet{app94} to
explain the orbital variability observed in a sample of eclipsing
variables and then later applied to the case of the black-widow pulsar
PSR B1957+20~\citep{app94}. The model has been then further extended
to Roche lobe filling systems like cataclysmic
variables~\citep{ric94}.  The model can be briefly summarized as
follows. If the donor star has internal deformations, then the
gravitational potential outside of the active star is (terms higher
than quadrupolar are ignored here):
\begin{equation}\label{eq:pot}
\phi(x)=-\frac{GM}{r}-\frac{3}{2}GQ_{ik}\frac{x_ix_k}{r^5}
\end{equation}
where $x_i$ and $x_k$ are Cartesian coordinates
measured from the center of mass of the star and $Q_{ik}$
is the quadrupole tensor (related to inertia tensor). 
For simplicity one can assume a circular orbit (as is legitimate to do in a binary like \saxj\,),
 an alignment between the donor spin axis and orbital angular momentum, and
that the stellar spin and orbit are synchronized.
If the Cartesian system is chosen so that the z-axis is the angular momentum axis and
the x-axis points from the center of mass to the companion star, then 
 $Q_{ik}$ reduces to $Q_{xx}=Q$. 
In a circular orbit, the relative velocity can be written as: 
\begin{equation}
v^2 = r\frac{d\phi}{dt}
\end{equation}
Therefore from Eq.~(\ref{eq:pot}) one can see that the relative velocity
$v$ is related to the time varying quadrupole $Q$. Since $v$ is also
related to the orbital period of the binary, it is clear that a time
varying mass quadrupole term induces a variation of the orbital
period. \citet{app92} suggests that the cause of the time varying
quadrupole $Q$ might be related to magnetic cycles of period $P_{\rm\,mod}$
(in a way analogous to the familiar 11 years-long magnetic solar cycles).
During these cycles, the magnetic field induces a redistribution of
the angular momentum in different layers of the star and allows a
transition between different equilibrium configurations.  The strength
of the surface magnetic field $B$ required to explain a variation
$\Delta P$ of the orbital period can be written as:
\begin{equation}
B^2{\sim}10\frac{GM^2}{R^4}\left(\frac{A}{R}\right)^2\frac{\Delta P}{P_{\rm mod}}
\end{equation}
The transport of angular momentum inside the star needs of course some
energy which \citet{app92} suggests might come either from the donor
internal nuclear burning reservoir or from tidal
heating~\citep{app94}.  This is in essence the Applegate model which
has been propose as a viable way to explain the behavior of many
binary systems. It seems therefore natural to extend it to the case of
LMXBs. For the case of \saxj\, one needs to assume a value for $P_{\rm
  mod}$ since the $T_{\rm asc}$ variations observed so far do not show
a complete cycle. By assuming $P_{\rm mod}=50\rm\,yr$,
$R=0.1\,R_{\odot}$, $M=0.07\msun$, $A=7\times10^{10}\rm\,cm$ one finds
$B\sim 1\rm\,kG$. \\

There are, however, two main problems with the Applegate model applied
to \saxj\, and to many other compact binaries considered here.  The
first is that the required B field is of the order of $1\rm\,kG$ which
is much larger than typical values thought to be present in fully convective stars.
However, some isolated low mass stars and brown dwarfs have been
observed with relatively strong surface B fields~\citep{mor10} and in
some cases with fields larger than 1 kG~\citep{rei12}.  Furthermore,
recent studies have provided observational evidence for the presence
of magnetic activity in at least four fully convective stars,
suggesting that the dynamo mechanism that produces stellar magnetic
fields operates also through convection despite the absence of the
tachocline, which is the boundary layer between radiative and
convective envelope where the magnetic fields are
generated~\citep{wri16}. 

The second problem, which seems more difficult to circumvent, was
discussed in a critical review by \citet{bri06}, who found that for
very low mass stars like NN-Serpentis (which is a non-accreting
post-common envelope binary with a 0.15$\msun$ companion and an
orbital period of 3 hours), the internal energy budget of the donor
star might be insufficient to generate the required donor distortion.
Even if one invokes the tidal heating mechanism proposed in
\citet{app94} there would still be insufficient energy available to
generate the required stellar distortions (see e.g. \citealt{bur15}).
As was the case for the mass-loss model, the only source of energy
left is the spin-down energy of the pulsar. In this case there needs
to be a viable mechanism to transport energy deeper in the donor star
which is able to generate a varying mass quadrupole. As noted by
\citet{app92}, if the donor star becomes more oblate then the mass
quadrupole $\Delta\,Q>0$ and the orbital period decreases.  The
opposite happens if $\Delta\,Q<0$: the orbital period increases.  The
observed behavior of most binaries considered in this work might then
be explained if, for some reason, some of them (like \saxj\, and other
diverging binaries) have $\Delta\,Q<0$ whereas all other converging
binaries have $\Delta\,Q>0$.  This idea remains highly speculative at
the moment since the problem of what happens in the deep layers of
irradiated stars has not been investigated yet.

\subsubsection{Enhanced Magnetic Braking}

This model could explain the sign and strength of $\dot{P}_b$ in
\saxj\, only if the donor magnetic field is sufficiently
strong. Indeed, the mass lost by the donor cannot be larger than the
one estimated in Section~\ref{sec:ml} since the energy budget does not
allow it. As an example, we follow the recipe provided by \citet{jus06}, in
which case the angular momentum lost by the binary
via magnetic braking is:
\begin{equation}\label{eq:mb}
\dot{J}_{\rm mb}=-\Omega_d\,B_s\,R_c^{13/4}\dot{M}_w^{1/2}(G\,M_c)^{-1/4}
\end{equation}
where $\Omega_d$ is the angular rotational frequency of the donor
star, $B_s$ is its dipolar magnetic field at the surface
and $\dot{M}_w$ is the amount of wind loss rate.  If we assume that
\saxj\, is changing orbital parameters mainly because of angular
momentum loss, then by rearranging Eq.(\ref{eq:orb}) we obtain:
\begin{equation}
 \dot{J}=\frac{\dot{a}}{a^{1/2}}M_{\rm NS}M_{\rm c}\left(\frac{G}{M}\right)^{1/2}\approx 2\times10^{35}\rm\,g\,cm^{2}s^{-2}
\end{equation}
Assuming that the donor is tidally locked, then Eq.(\ref{eq:mb}) gives
the strength of the minimum $B_s$ field required which is of the order of
$10^3$--$10^4$ G for a maximum wind loss rate of
$10^{-9}\rm\msun\rm\,yr^{-1}$, even stronger than the value
calculated for the Applegate model.  This shows that the magnetic braking model is
unlikely to be the correct one.  Furthermore, such explanation cannot
work in several other binaries since at least in some neutron star
LMXBs the sign of the observed orbital period derivative is opposite
to that expected when magnetic braking is the main driver of binary evolution.

\section{Conclusions}

We have studied the AMXP \saxj\, at radio-wavelengths during
quiescence in 2014 and during its last 2015 outburst.  We have not
detected radio pulsations and we place strong constraints on the flux
density of the putative radio pulsar which, if really active, needs to
be either among the 10\% dimmest pulsars known or fully obscured by
radio absorbing material (which would also be atypical at the
relatively high radio observing frequencies we searched at).  The
study of the orbital evolution of the system has been extended to
include the 2015 outburst, and we find two possible interpretations of
the data: either the orbit is expanding with stochastic fluctuations
around the mean or the system is shrinking with a change of sign
around 2011.

In the first case the pulsar spin-down power is ablating the companion
with an efficiency for the conversion of impinging power to mass-loss
of the order of 40\%.

Alternatively the Applegate model can explain the behavior of \saxj\,
if a strong surface magnetic field of the order of a kG is
present. The source of energy that powers this field needs to be the
spin-down power of the pulsar but there is no evidence that such large
fields exist in the donor star of \saxj\, or that they can be
generated by the pulsar wind/high energy irradiation.  This requires
further theoretical investigation.


\bigskip

\acknowledgements{ We would like to thank M. Fortin for providing the
  data for the mass-radius relations of several neutron star models.
  We would like to thank E.P.J. van den Heuvel for interesting
  discussions and suggestions.  A.P. acknowledges support from an NWO
  Vidi fellowship. R.W. was supported by an NWO Top Grant, Module 1.
  J.W.T.H. is an NWO Vidi Fellow.  A.J. and J.W.T.H. acknowledge
  funding from the European Research Council under the European
  Union's Seventh Framework Programme (FP7/2007-2013) / ERC grant
  agreement nr. 337062 (DRAGNET). Some of the results presented in
  this paper were based on observations obtained with GBT+GUPPI. We
  would therefore like to express our gratitude towards the National
  Radio Astronomy Observatory (NRAO) - a facility of the United States
  National Science Foundation (NSF) - responsible for operating the
  Green Bank Telescope.  Computational support for radio data analysis
  was provided by supercomputer Cartesius - a service offered by the
  Dutch SURFsara. Part of the scientific results reported in this
  article are based the observations made by the \textit{Chandra X-ray
  Observatory}. This research has made use of software provided by the
  Chandra X-ray Center (CXC) in the application packages CIAO. }


\end{document}